\documentclass[%
 reprint,
 amsmath,amssymb,
 aps,
]{revtex4-2}

\usepackage{graphicx}
\usepackage{dcolumn}
\usepackage{bm}

\begin{document}

\preprint{APS/123-QED}

\title{Spin-wave assisted synchronization in 2D arrays of spin torque oscillators}

\author{Fangzhou Ai}
\author{Vitaliy Lomakin}%
 \email{vlomakin@ucsd.edu}
\altaffiliation[Also at ]{Program in Materials Science and Engineering, University of California, San Diego, La Jolla 92093, U.S.A.}
\affiliation{%
 Department of Electrical and Computer Engineering, University of California San Diego, La Jolla 92093, U.S.A
}%
\affiliation{%
 Center for Memory and Recording Research, University of California San Diego, La Jolla 92093, U.S.A
}%


\begin{abstract}

This work reports on achieving synchronization in a large 2D array of spin transfer torque nano-oscillators (STNOs) by means of propagating spin waves. The synchronization is made possible by means of a honeycomb structure in the magnetic film with interchanging regions with higher and lower magnetic damping. Such a structure creates a network of nearest neighbor interactions between the STNOs with an equal properly tuned phase shift. Micromagnetic simulations demonstrate efficiency of synchronization, including phase and frequency locking across the entire large STNO array. Synchronization is obtained at zero and room temperatures as well as for structure with distributions of the material parameters. Synchronization can be tuned by enabling or disabling next-nearest neighbor interactions and such tunability allows selectively synchronize or de-synchronize sub-arrays of the entire large array. The presented ideas may find applications as microwave sources and for neuromorphic computing and signal processing.

\end{abstract}

\maketitle


\section{Introduction}
Spin-transfer nano-oscillators (STNOs) are non-linear devices that convert a direct current into a microwave voltage or current output~\cite{4802339, 1420682, 6832877} that can be tuned in a broach range of frequencies. STNOs gained a significant attention due to their unique physics of operation and potential applications, such as microwave communications~\cite{doi:10.1126/sciadv.adk1430, Choi2014}, signal processing~\cite{Litvinenko2020, 8007619}, and neuromorphic computing~\cite{GONZALEZ2024101173, 8734967, Böhnert2023, PhysRevApplied.12.024052, Romera2018}. A challenge is synchronizing a large number of STNOs~\cite{PhysRevLett.95.067203, 10192065, Tsunegi2018}. Synchronization, in principle, can be achieved by means of magnetostatic coupling as well as evanescent and propagating spin wave coupling for STNOs defined as point contacts in a magnetic film~\cite{kumar2023mutualsynchronizationspintorque, PhysRevB.74.104401}. Magnetostatic or evanescent spin wave coupling mechanisms do not include propagation related phase shifts and thus are easier to achieve but they also lack some of the control flexibility associated with the phase propagation. While models of mutual synchronization between two STNOs by means of propagating spin waves are established~\cite{mancoff2005phase, kaka2005mutual}, synchronization more STNOs remains a challenge. The inherent non-linearity and complicated interactions between a large number of STNOs result in rich and complex dynamics that is difficult to account for~\cite{PhysRevB.74.104401}. Despite some analysis towards special structures~\cite{10192065, Castro2022}, synchronization of a large 2D array of STNOs by means of propagating spin waves is a challenge. 

Here, we present an approach to achieve synchronization in a large 2D array of point-contact STNOs by means of propagating spin waves in a magnetic film and demonstrate the synchronization via micromagnetic simulations. We introduce a honeycomb pattern consisting of a low-damping and high-damping regions (Fig. 1). The main ideas are to create equal distances between the STNOs to result in an in-phase spin interactions and to allow only near-neighbor interactions between the STNOs leading to their collective behavior. This approach overcomes the challenges posed by the complicated interactions in large STNO arrays and provides a pathway towards achieving STNO synchronization.

\section{Methodology and results}
We consider synchronization in arrays of $N$ STNOs formed in a $d=3$nm thick soft magnetic film with the saturation magnetization $M_s=6.37\times 10^{-5}$A/m and exchange constant $A_{ex}=1.4\times 10^{-11}$J/m with no anisotropy (Fig. 1). The magnetization is initialized along the vertical ($+z$) direction and is relaxed to its equilibrium state under an external perpendicular applied field $\mathbf{H}_{app}$. STNOs are formed by STT set via a point contact of $r_{STNO}=50$nm radius with the current of $\mathbf{I}_{stt}$ with the $z$-directed polarization with the polarization efficiency of $p=0.4$.

The vertical applied field sets the STNO frequency and leads to magnetostatic forward volume wave~\cite{prabhakar2009spin} spin waves that are isotropic with respect to the in-plane propagation and whose lowest mode wavelength is given by the following dispersion relation~\cite{prabhakar2009spin}:
\begin{equation} \label{eq:msfvw}
    \omega=\omega_{ex}\left[ \omega_{ex}+\omega_M\left(1-\frac{1-e^{-k_{\perp}d}}{k_{\perp}d}\right)\right],
\end{equation}
where $\omega$ is the operation frequency, $\omega_M=\gamma\mu_0M_s$, $\omega_{ex}=\gamma\mu_0(|\mathbf{H}_{app}|-M_s)+2\gamma A_{ex}k_{\perp}^2/M_s$, $\gamma$ is the gyro-magnetic ratio, $\mu_0$ is the vacuum permeability and $k_{\perp}$ is the spin wave wavenumber. For the chosen thin film, only the lowest spin wave mode exists~\cite{prabhakar2009spin}, i.e., the $z$-direction wavenumber component $k_z=0$. The STNOs are driven by STT currents, which ensures their operation with a large-angle precession, leading to equal-strength interactions between all STNO pairs.
The synchronization is manifested via frequency and phase locking of the STNO auto-oscillations, i.e., a stable phase difference between any two oscillators at the same frequency of precession. To quantify the synchronization, let $\mathbf{m}_i(\theta_i,\phi_i,t)$ be the spatial averaged magnetization of the $i^{th}$ STNO, $\theta_i$ be the corresponding azimuthal angle, and $\phi_i$ be the elevation angle. The phase difference can be expressed as $\Delta_{i,j}(t)=\cos^{-1}[\mathbf{m}_i(\theta_i,\phi_i,t), \mathbf{m}_j(\theta_j,\phi_j,t)]$, and if it is independent of time, then the $i^{th}$ and $j^{th}$ STNOs are phase locked. If this holds for all pairs of STNOs, then global synchronization is achieved. The case of all $\Delta_{i,j}=0$ corresponds to a global in-phase synchronization. A measure of synchronization is the output power averaged over a time period $T_0$:
\begin{equation}
    P=\frac{1}{T_0}\int_0^{T_0}(\sum_{i=1}^Nm_{ix})^2dt.
\end{equation}
Such an output power corresponds to the total current generated by the array if an in-plane layer is added to each STNO to generate an electric signal via magnetoresistance. For a non-synchronized array in which a single STNO produced power $P_0=\int_0^{T_0}(m_x)^2dt/T_0$, the total power is $P=NP_0$ because of the arbitrary phase differences, whereas for a perfectly synchronized array, $P_s=N^2P_0$, where $P_s$ is the maximal achievable total power. Based on the total output power scaling we can define the synchronization efficiency as $\eta=P/P_s \leq 1$.
\begin{figure*} \label{fig:p1}
    \includegraphics[width=16cm]{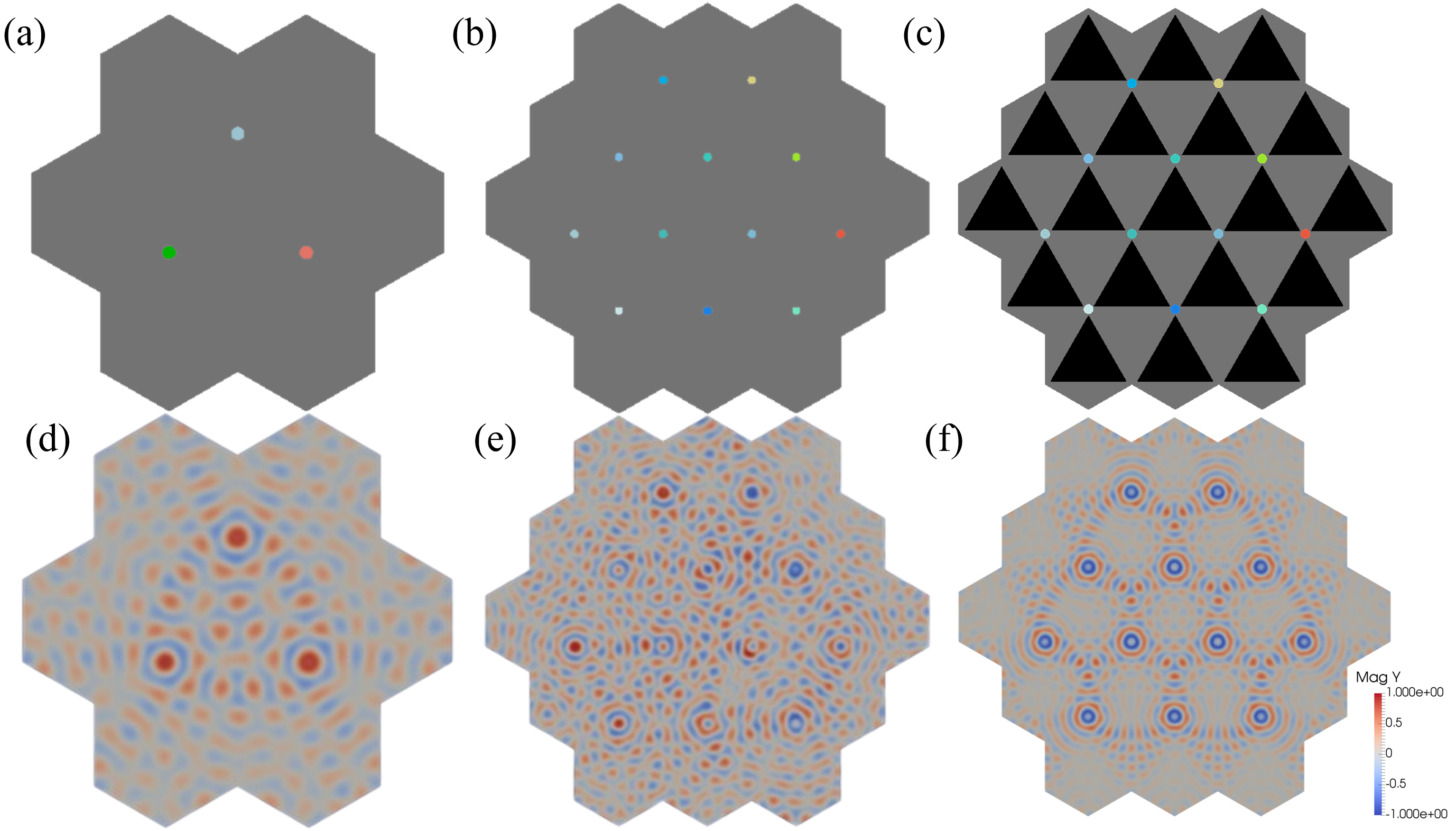}
    \caption{(a) and (b) Arrays of 3 and 12 STNO in a uniform magnetic film with snapshots of the corresponding $m_y$ magnetization component in (d) and (e). (c) Array of 12 STNO is a honeycomb pattern magnetic film with regions of lower (grey) and higher (black) magnetic damping, and a snapshot of the corresponding $m_y$ magnetization component in (f). Synchronization is not obtained for the 12 STNO case with a homogeneous film but is obtained in the honeycomb pattern film.}
\end{figure*}

Synchronization between a pair of STNOs by means of spin waves can be readily achieved~\cite{mancoff2005phase, kaka2005mutual}. However, synchronizing a large 2D array of STNOs by means of spin waves is challenging. As a larger array, we first consider synchronization in an array of $N=3$ STNOs arranged at vertices of a triangle (Fig. 1a, d). The applied field and current are chosen as $\mathbf{H}_{app}=1.32$ T$\hat{\mathbf{z}}$ and $\mathbf{I}_{stt}=-22.5$ mA$\hat{\mathbf{z}}$, respectively. Via Eq.~\eqref{eq:msfvw}, these parameters correspond to the STNO oscillating frequency $\omega/2\pi\approx30$ GHz with spin wave wavelength of $\lambda\approx60$ nm, and the elevation angle of the STNO precession is $\theta_e\approx30^{\circ}$. The separation between the STNOs is chosen as $d_{ee}=420$ nm$=7\lambda$ such that it corresponds to an integer number of spin wave wavelengths, so that in-phase synchronization is obtained. Global synchronization is achieved in this case, similar to the case of two STNOs (Fig. 1d). We then extend the structure to a hexagonal array of   STNOs (Fig. 1b, e). In this case, a chaotic behavior is obtained with no synchronization (Fig. 1e). Trying to move the STNO locations or making arranging them in a different, e.g., square, array still leads to a similar chaotic behavior. This lack of synchronization is attributed to the fact that the spin waves generated by each STNO interact with other STNOs at different phases, thus leading to hectic, out of phase interaction patterns. The interactions carry varying phase differences originating from neighbor STNOs at different distances. Furthermore, the strength of these interactions also varies with the distance, contributing to non-uniform effects across the array. These factors result in accumulated discrepancies that significantly disrupt global synchronization. The interplay of inconsistent phase shifts and variable interaction strengths results in a complex dynamic, where the collective behavior deviates from the expected synchronized order, leading to the chaotic patterns observed in larger arrays. 

In exploring the synchronization pattern further with various geometry configurations, we find that global synchronization is determined by the presence of next-nearest-neighbor and further interactions and possibly unequal separations between STNOs. An approach to limit the chaotic behavior caused by the long-range interactions is to block/absorb the spin waves. This understanding leads to introducing the honeycomb structure in Fig. 1c. In this structure, the STNOs are positioned in a hexagonal pattern in a magnetic film. The magnetic properties of the film, except the damping constant, are homogeneous across the film. The damping constant on the other hand has triangular regions of high values (black color in Fig. 1c) and low values (grey color in Fig. 1c). The low- and high-damping values can be either determined by ion irradiation~\cite{FASSBENDER2008579} or by defining additional nano-contacts with polarized current, which would either increase or decrease the effective damping constant as set by the strength and direction of the STT current~\cite{PhysRevLett.123.257201}. The large and low damping regions have short- and long-range spin wave propagation, respectively. The size of the neck (formed by the high damping regions) $d_{neck}$ is comparable to the spin wave wavelength $\lambda$, effectively preventing a strong spin wave transmission through the neck, such that the spin wave excitation in the neck is determined by the corresponding STNO. Additionally, the hexagonal array ensures that the nearest-neighbor distances are the same, which leads to the array symmetry with equal spin phase propagation phases enabling global synchronization. In this particular structure, $d_{neck}=80$ nm$\approx1.33\lambda$, the higher damping constant is $\alpha_{high}=0.2$ and lower damping constant is $\alpha_{low}=0.01$. With this high-/low-damping structure, we indeed obtain a global synchronization for the same array of $N=12$ STNOs, as is evident via the high synchronization efficiency in Fig. 1f.

\begin{figure} \label{fig:p2}
    \includegraphics[width=7cm]{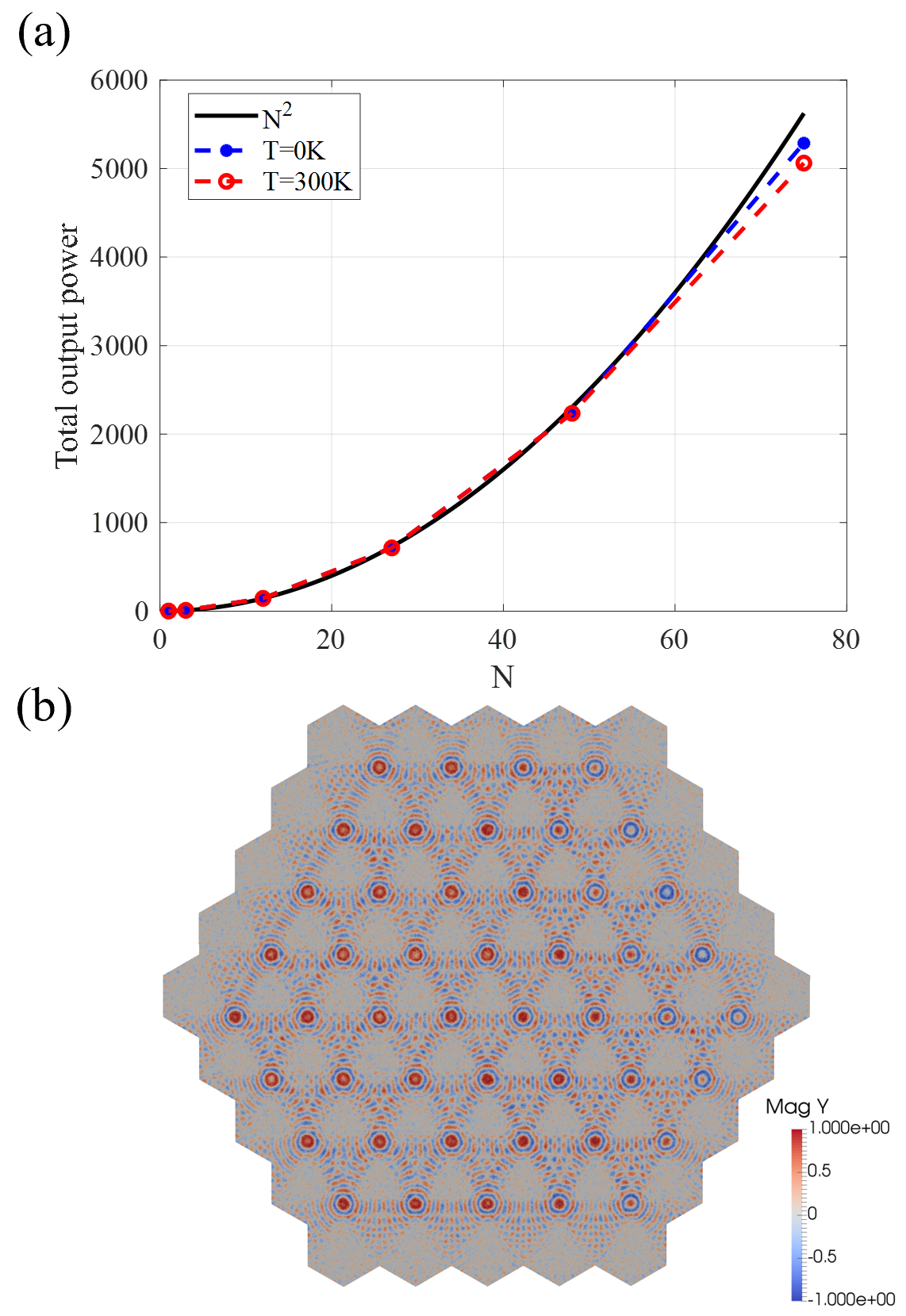}
    \caption{(a) Total power at $T=0$ and 300K vs. the theoretical maximum; (b) a snapshot of $m_y$ for $T=300K$ for 48 STNOs.}
\end{figure}

Achieving high-efficiency synchronization requires finding proper structure parameters. To find proper combination of $\mathbf{H}_{app}$ and $\mathbf{I}_{stt}$, we first choose a set of $\mathbf{I}_{stt}$ from 22.0 mA to 23.0 mA and sweep the $\mathbf{H}_{app}$ from 1 T$\hat{\mathbf{z}}$ to 2 T$\hat{\mathbf{z}}$ with 1000 ns duration. We calculate the synchronization efficiency $\eta$ by integrating all the $m_{ix},i\in[1,N]$ with a time period $T_0=1$ ns, and find under what $\mathbf{H}_{app}$ and $\mathbf{I}_{stt}$ the best synchronization efficiency $\eta$ is obtained. The resulting total output power is averaged over the last 10 ns in each simulation.
We consider STNO arrays of different sizes, up to 48 STNOs and perform simulations at zero and room  (300 K) temperatures with all STNOs driven by the same $\mathbf{H}_{app}$ and $\mathbf{I}_{stt}$ (Fig. 2a). Global synchronization is achieved with a high synchronization efficiency $\eta$ in all these cases. Synchronization is also manifested through a reduced linewidth. For a single STNO, the linewidth at full width at half maximum is close to 1.8 GHz. For the synchronized arrays, the linewidth is below 0.2 GHz even for the room temperature cases, which is much smaller. Figure 2b further shows the spin wave and STNOs phase patterns with $N=48$ at the temperature of 300K when global synchronization is present. We find only a small phase shift across the STNOs in the array, which is due to thermal fluctuation and minor mismatches between $d_{ee}$ and $\lambda$, resulting in slight drop of $\eta$ compared to smaller arrays or zero temperature results. The results demonstrate the same spin wave interference patterns in all low damping regions and strong absorption at the edges of each high damping region. These consistent spin wave patterns provide evidence that the honeycomb design effectively limits a long-range spin wave propagation and achieves nearest-neighbor only interactions between the STNOs.

\begin{figure} \label{fig:p3}
    \includegraphics[width=7cm]{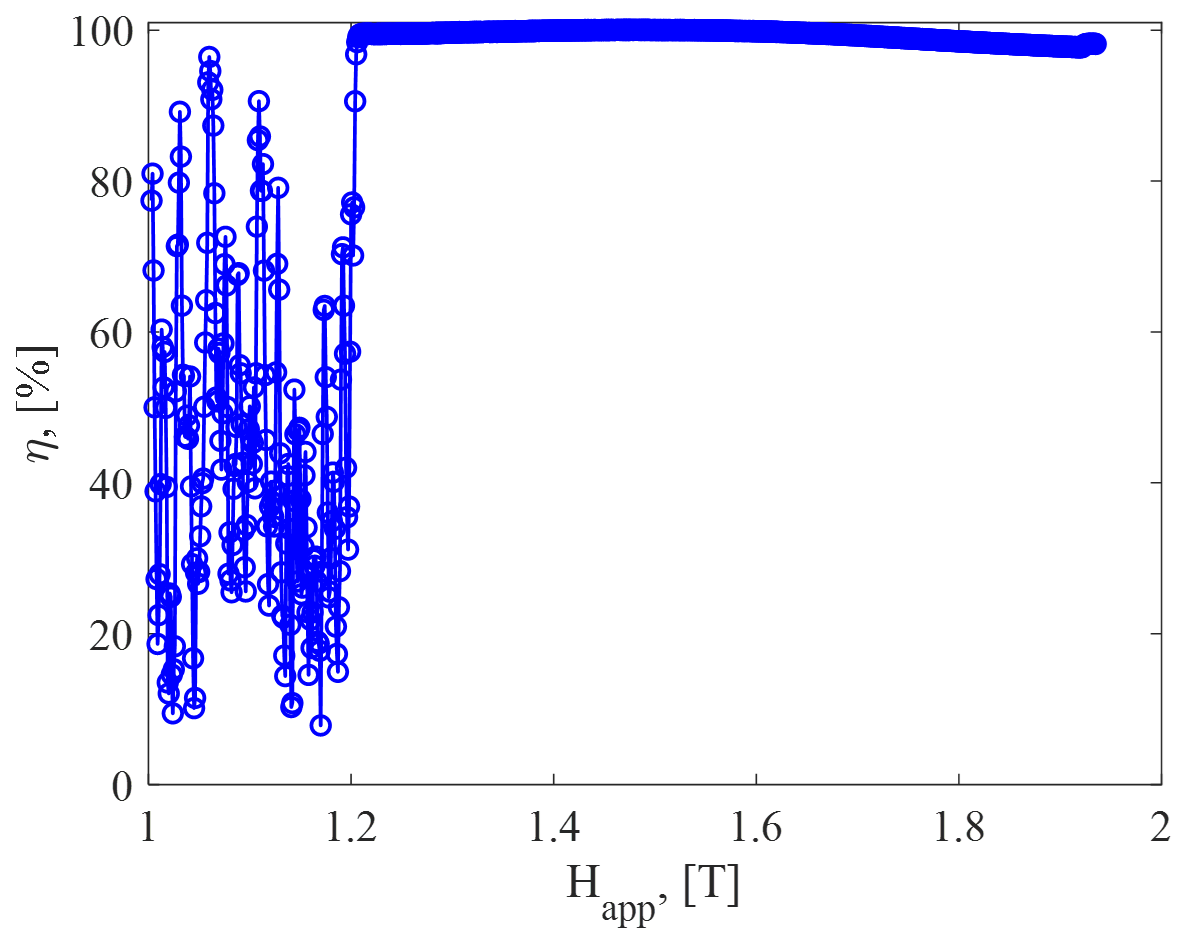}
    \caption{Efficiency $\eta$ for different $\mathbf{H}_{app}$.}
\end{figure}

We note that synchronization is achieved only in a certain parameter range. To characterize such a range, Fig. 3 shows $\eta$ for different STNO array sizes as a function of $\mathbf{H}_{app}$, keeping the rest of the parameters the same. We observe a sharp transition from a low and irregular efficiency, indicating a lack of synchronization, to a stable high efficiency for a proper range of $\mathbf{H}_{app}$. Smaller $N$ has a wider synchronization range whereas for larger $N$, only a narrow range of $\mathbf{H}_{app}$ and $\mathbf{I}_{stt}$ leads to synchronization. It can be concluded that the lack of synchronization for smaller $\mathbf{H}_{app}$ is due to a higher spin wave magnitude, so that the spin waves can pass through the neck barriers, and thus reach the next-nearest neighbors, leading to a chaotic behavior. With a stronger $\mathbf{H}_{app}$, the spin wave magnitude decreases preventing the next-nearest-neighbor interactions, resulting in synchronization. Too large $\mathbf{H}_{app}$ also may destroy synchronization, which is attributed to a weak spin wave excitation, resulting in too weak coupling between the STNOs. Other parameters, e.g., $\mathbf{I}_{stt}$, affect the synchronization range in a similar way, so that various ranges of parameters can be tuned. The synchronization range for very large arrays is narrower because the phase differences between the STNOs accumulates eventually resulting in destroying the global synchronization.

\begin{figure} \label{fig:p4}
    \includegraphics[width=7cm]{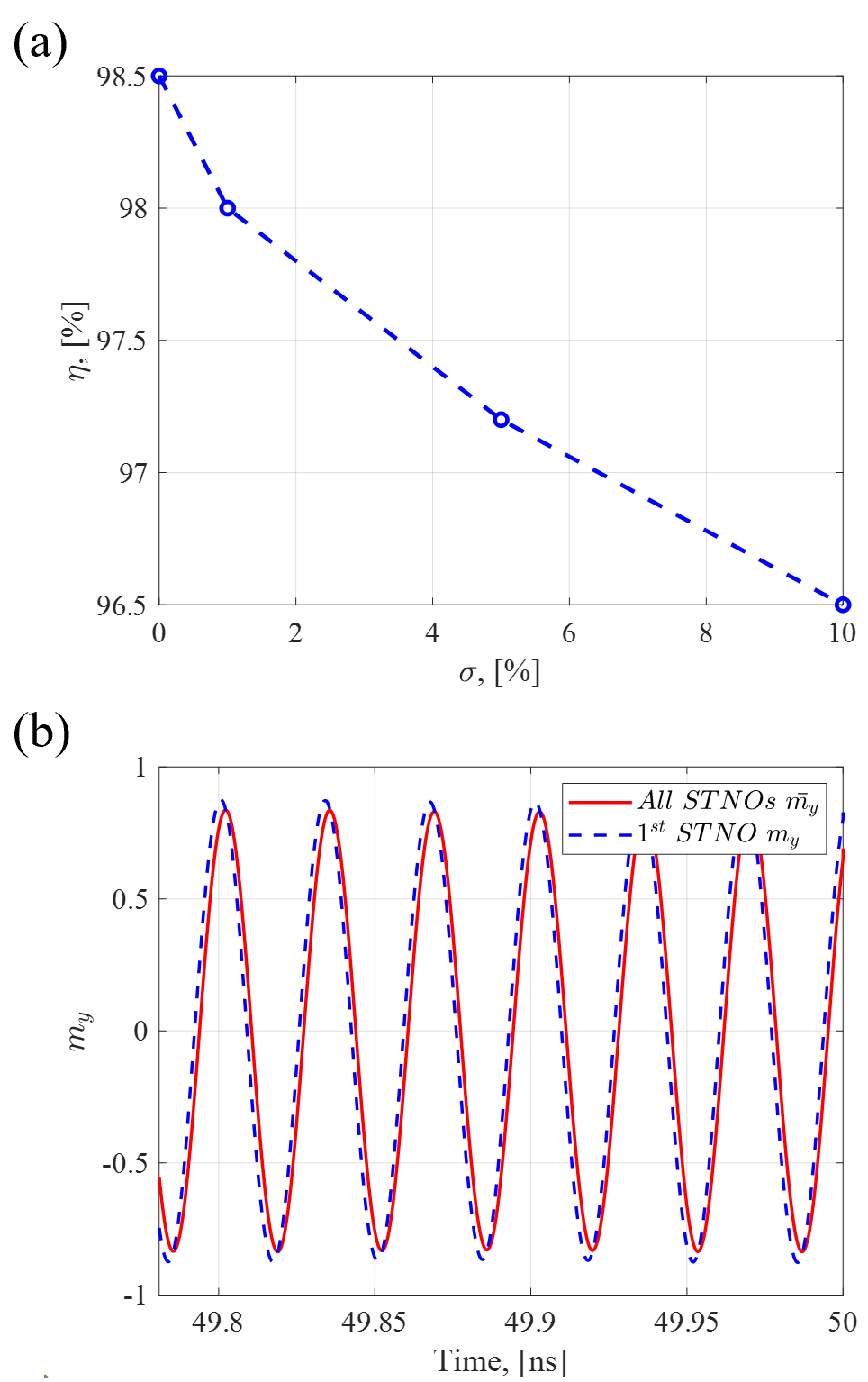}
    \caption{Synchronization vs. different uniform distributions with mean value $\mu=6.37\times 10^5$ A/m and deviation $\sigma=0,1,5,10\%$ for $M_s=\mu\times(1\pm\sigma)$ and $N=48$. (a) Efficiency $\eta$; (b) averaged $m_y$ for all the STNO vs. $m_y$ of the first STNO as a function of time for $\sigma=10\%$.}
\end{figure}

The synchronization range also is reduced at elevated temperatures due to stochastic (Brownian-like) effects as well as random distributions of the material properties. To characterize the effects of the parameter distributions, Fig. 4a shows the synchronization efficiency for an STNO array of $N=48$ with the saturation magnetization of uniform distribution with the mean value of $\mu=6.37\times 10^{5}$ A/m and deviations of $\sigma=0,1,5,10\%$, i.e. $M_s=\mu\times(1\pm\sigma/2)$. We find that $\eta$ slightly drops from 98.5\% to 96.5\%, which is still well above 90\%. Figure 4b exhibits a snapshot of comparison between $\bar{m}_y$, $m_y$ averaged over the 48 STNOs, and $m_y$ of the first STNO located at the top left corner of the array for the largest considered $\sigma$. We find that $\bar{m}_y$ overlaps with $m_y$, showing a robust global synchronization is still maintained. The minor decline in synchronization efficiency $\eta$ is mostly attributed to the resulting distribution in the STNO frequencies, which leads to extra phase shifts between the STNOs.

\begin{figure} \label{fig:p5}
    \includegraphics[width=8cm]{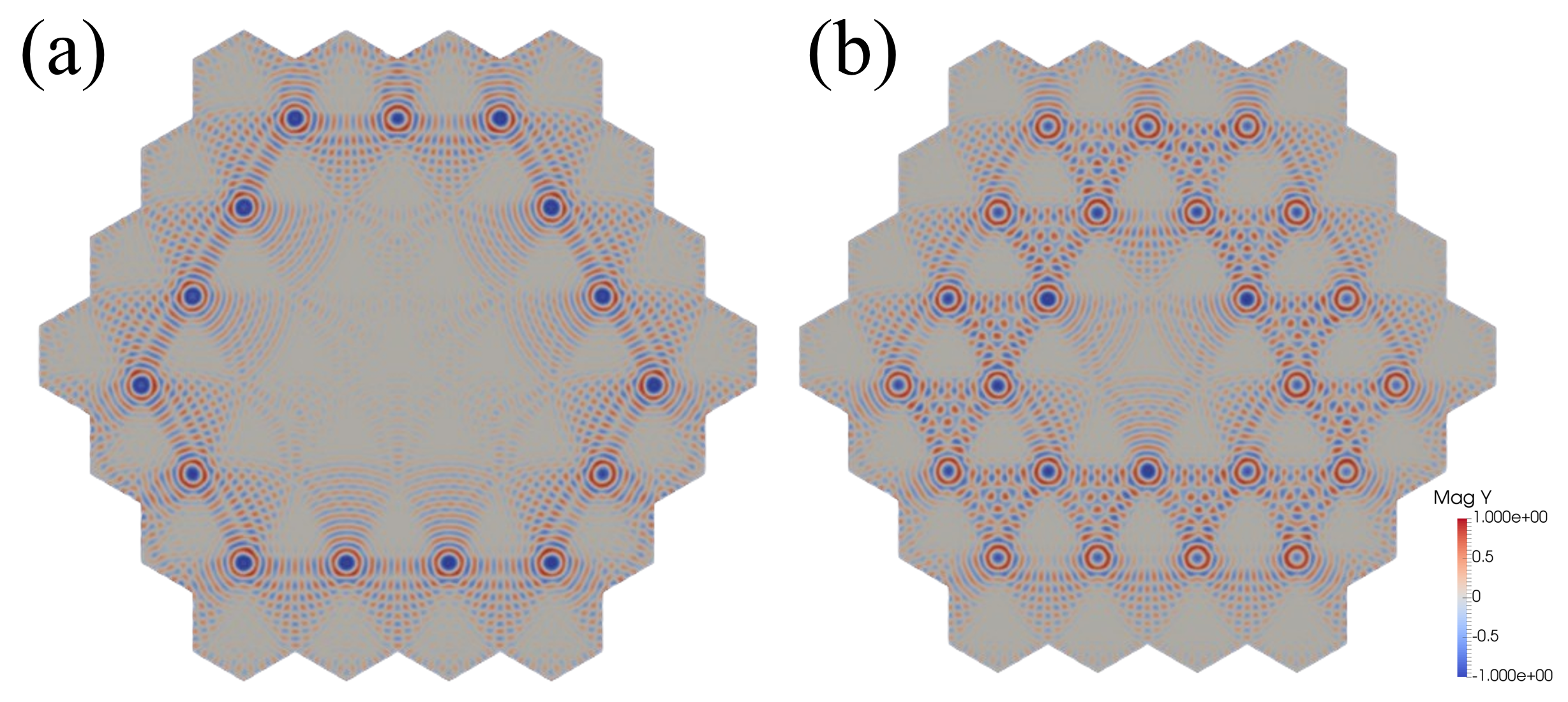}
    \caption{Magnetization along y-direction of synchronized (a) ring and (b) concentric loops. These two structures are built by turning off some STNOs on the model of size $N=27$.}
\end{figure}

The ideas leading to global synchronization also allow creating synchronization in subarrays of the entire large array, providing flexibility for various application requirements. For example, by turning on and off some STNOs at certain position by locally reducing or extending the spin wave propagation length, we can create a synchronized STNO ring, e.g., by only activating the outermost shell of the array (Fig. 5a) or concentric loops by activating the inner shells (Fig. 5b). 

\section{Conclusions}
In conclusion, synchronization of a large array of STNOs by propagating spin waves is possible when a mechanism restricting long range interactions between the STNOs is chosen together with properly tuned parameters. The mechanism chosen here was based on controlling the magnetic damping, which controls the spin wave propagation length. Adaptive control of the spin wave propagation length also allows defining synchronization in selected subarrays of the entire array with tuned paths of synchronization. Additional mechanisms may also be possible, e.g., a network of nanowires connecting the STNOs may result in a similar behavior, albeit without an easy potential of adaptive tuning offered by tuning the spin wave propagation length. Potentially, additional tuning mechanisms may also be introduced, e.g., if the current in each STNO is tuned separately, which would allow synchronization in very large STNO arrays. The presented ideas may find applications in creating microwave generators with increased output power as well as in neuromorphic computing and signal processing. In particular, synchronized subarrays of large arrays of STNOs demonstrate diverse synchronization patterns, akin to the dynamics observed in reservoir computing networks. These patterns can be leveraged to perform complex computational tasks, making them relevant for next-generation adaptive systems that require real-time processing capabilities~\cite{DIEBNER2002781}. The presented ideas can be extended to systems with alternative coupling mechanism, e.g., to arrays of STNOs or spin Hall oscillators coupled via elastic waves with the magnetoelastic effect or additional wave mechanisms in other physical systems.

\begin{acknowledgments}
This work was supported in part by the Quantum Materials for Energy Efficient Neuromorphic-Computing (Q-MEEN-C), an Energy Frontier Research Center funded by the U.S. Department of Energy, Office of Science, Basic Energy Sciences under Award No. DESC0019273. The work was also supported in part by Binational Science Foundation, grant \#2022346. The work used Purdue Anvil cluster at Rosen Center for Advanced Computing (RCAC) in Purdue University and Delta cluster at National Center for
Supercomputing Applications (NCSA) in University of Illinois Urbana-Champaign through allocation ASC200042 from the Advanced Cyberinfrastructure Coordination Ecosystem: Services \& Support (ACCESS) program ~\cite{10.1145/3569951.3597559}, which is supported by National Science Foundation grants \#2138259, \#2138286, \#2138307, \#2137603, and \#2138296.
\end{acknowledgments}




\bibliography{apssamp}

\begin{thebibliography}{25}%
\makeatletter
\providecommand \@ifxundefined [1]{%
 \@ifx{#1\undefined}
}%
\providecommand \@ifnum [1]{%
 \ifnum #1\expandafter \@firstoftwo
 \else \expandafter \@secondoftwo
 \fi
}%
\providecommand \@ifx [1]{%
 \ifx #1\expandafter \@firstoftwo
 \else \expandafter \@secondoftwo
 \fi
}%
\providecommand \natexlab [1]{#1}%
\providecommand \enquote  [1]{``#1''}%
\providecommand \bibnamefont  [1]{#1}%
\providecommand \bibfnamefont [1]{#1}%
\providecommand \citenamefont [1]{#1}%
\providecommand \href@noop [0]{\@secondoftwo}%
\providecommand \href [0]{\begingroup \@sanitize@url \@href}%
\providecommand \@href[1]{\@@startlink{#1}\@@href}%
\providecommand \@@href[1]{\endgroup#1\@@endlink}%
\providecommand \@sanitize@url [0]{\catcode `\\12\catcode `\$12\catcode `\&12\catcode `\#12\catcode `\^12\catcode `\_12\catcode `\%12\relax}%
\providecommand \@@startlink[1]{}%
\providecommand \@@endlink[0]{}%
\providecommand \url  [0]{\begingroup\@sanitize@url \@url }%
\providecommand \@url [1]{\endgroup\@href {#1}{\urlprefix }}%
\providecommand \urlprefix  [0]{URL }%
\providecommand \Eprint [0]{\href }%
\providecommand \doibase [0]{https://doi.org/}%
\providecommand \selectlanguage [0]{\@gobble}%
\providecommand \bibinfo  [0]{\@secondoftwo}%
\providecommand \bibfield  [0]{\@secondoftwo}%
\providecommand \translation [1]{[#1]}%
\providecommand \BibitemOpen [0]{}%
\providecommand \bibitemStop [0]{}%
\providecommand \bibitemNoStop [0]{.\EOS\space}%
\providecommand \EOS [0]{\spacefactor3000\relax}%
\providecommand \BibitemShut  [1]{\csname bibitem#1\endcsname}%
\let\auto@bib@innerbib\@empty
\bibitem [{\citenamefont {Slavin}\ and\ \citenamefont {Tiberkevich}(2009)}]{4802339}%
  \BibitemOpen
  \bibfield  {author} {\bibinfo {author} {\bibfnamefont {A.}~\bibnamefont {Slavin}}\ and\ \bibinfo {author} {\bibfnamefont {V.}~\bibnamefont {Tiberkevich}},\ }\bibfield  {title} {\bibinfo {title} {Nonlinear auto-oscillator theory of microwave generation by spin-polarized current},\ }\href {https://doi.org/10.1109/TMAG.2008.2009935} {\bibfield  {journal} {\bibinfo  {journal} {IEEE Transactions on Magnetics}\ }\textbf {\bibinfo {volume} {45}},\ \bibinfo {pages} {1875} (\bibinfo {year} {2009})}\BibitemShut {NoStop}%
\bibitem [{\citenamefont {Slavini}\ and\ \citenamefont {Kabos}(2005)}]{1420682}%
  \BibitemOpen
  \bibfield  {author} {\bibinfo {author} {\bibfnamefont {A.}~\bibnamefont {Slavini}}\ and\ \bibinfo {author} {\bibfnamefont {P.}~\bibnamefont {Kabos}},\ }\bibfield  {title} {\bibinfo {title} {Approximate theory of microwave generation in a current-driven magnetic nanocontact magnetized in an arbitrary direction},\ }\href {https://doi.org/10.1109/TMAG.2005.845915} {\bibfield  {journal} {\bibinfo  {journal} {IEEE Transactions on Magnetics}\ }\textbf {\bibinfo {volume} {41}},\ \bibinfo {pages} {1264} (\bibinfo {year} {2005})}\BibitemShut {NoStop}%
\bibitem [{\citenamefont {Dumas}\ \emph {et~al.}(2014)\citenamefont {Dumas}, \citenamefont {Sani}, \citenamefont {Mohseni}, \citenamefont {Iacocca}, \citenamefont {Pogoryelov}, \citenamefont {Muduli}, \citenamefont {Chung}, \citenamefont {Dürrenfeld},\ and\ \citenamefont {Åkerman}}]{6832877}%
  \BibitemOpen
  \bibfield  {author} {\bibinfo {author} {\bibfnamefont {R.~K.}\ \bibnamefont {Dumas}}, \bibinfo {author} {\bibfnamefont {S.~R.}\ \bibnamefont {Sani}}, \bibinfo {author} {\bibfnamefont {S.~M.}\ \bibnamefont {Mohseni}}, \bibinfo {author} {\bibfnamefont {E.}~\bibnamefont {Iacocca}}, \bibinfo {author} {\bibfnamefont {Y.}~\bibnamefont {Pogoryelov}}, \bibinfo {author} {\bibfnamefont {P.~K.}\ \bibnamefont {Muduli}}, \bibinfo {author} {\bibfnamefont {S.}~\bibnamefont {Chung}}, \bibinfo {author} {\bibfnamefont {P.}~\bibnamefont {Dürrenfeld}},\ and\ \bibinfo {author} {\bibfnamefont {J.}~\bibnamefont {Åkerman}},\ }\bibfield  {title} {\bibinfo {title} {Recent advances in nanocontact spin-torque oscillators},\ }\href {https://doi.org/10.1109/TMAG.2014.2305762} {\bibfield  {journal} {\bibinfo  {journal} {IEEE Transactions on Magnetics}\ }\textbf {\bibinfo {volume} {50}},\ \bibinfo {pages} {1} (\bibinfo {year} {2014})}\BibitemShut {NoStop}%
\bibitem [{\citenamefont {Hamadeh}\ \emph {et~al.}(2023)\citenamefont {Hamadeh}, \citenamefont {Slobodianiuk}, \citenamefont {Moukhader}, \citenamefont {Melkov}, \citenamefont {Borynskyi}, \citenamefont {Mohseni}, \citenamefont {Finocchio}, \citenamefont {Lomakin}, \citenamefont {Verba}, \citenamefont {de~Loubens}, \citenamefont {Pirro},\ and\ \citenamefont {Klein}}]{doi:10.1126/sciadv.adk1430}%
  \BibitemOpen
  \bibfield  {author} {\bibinfo {author} {\bibfnamefont {A.~A.}\ \bibnamefont {Hamadeh}}, \bibinfo {author} {\bibfnamefont {D.}~\bibnamefont {Slobodianiuk}}, \bibinfo {author} {\bibfnamefont {R.}~\bibnamefont {Moukhader}}, \bibinfo {author} {\bibfnamefont {G.}~\bibnamefont {Melkov}}, \bibinfo {author} {\bibfnamefont {V.}~\bibnamefont {Borynskyi}}, \bibinfo {author} {\bibfnamefont {M.}~\bibnamefont {Mohseni}}, \bibinfo {author} {\bibfnamefont {G.}~\bibnamefont {Finocchio}}, \bibinfo {author} {\bibfnamefont {V.}~\bibnamefont {Lomakin}}, \bibinfo {author} {\bibfnamefont {R.}~\bibnamefont {Verba}}, \bibinfo {author} {\bibfnamefont {G.}~\bibnamefont {de~Loubens}}, \bibinfo {author} {\bibfnamefont {P.}~\bibnamefont {Pirro}},\ and\ \bibinfo {author} {\bibfnamefont {O.}~\bibnamefont {Klein}},\ }\bibfield  {title} {\bibinfo {title} {Simultaneous multitone microwave emission by dc-driven spintronic nano-element},\ }\href {https://doi.org/10.1126/sciadv.adk1430} {\bibfield  {journal} {\bibinfo  {journal} {Science
  Advances}\ }\textbf {\bibinfo {volume} {9}},\ \bibinfo {pages} {eadk1430} (\bibinfo {year} {2023})},\ \Eprint {https://arxiv.org/abs/https://www.science.org/doi/pdf/10.1126/sciadv.adk1430} {https://www.science.org/doi/pdf/10.1126/sciadv.adk1430} \BibitemShut {NoStop}%
\bibitem [{\citenamefont {Choi}\ \emph {et~al.}(2014)\citenamefont {Choi}, \citenamefont {Kang}, \citenamefont {Cho}, \citenamefont {Oh}, \citenamefont {Shin}, \citenamefont {Park}, \citenamefont {Jang}, \citenamefont {Min}, \citenamefont {Kim}, \citenamefont {Park},\ and\ \citenamefont {Park}}]{Choi2014}%
  \BibitemOpen
  \bibfield  {author} {\bibinfo {author} {\bibfnamefont {H.~S.}\ \bibnamefont {Choi}}, \bibinfo {author} {\bibfnamefont {S.~Y.}\ \bibnamefont {Kang}}, \bibinfo {author} {\bibfnamefont {S.~J.}\ \bibnamefont {Cho}}, \bibinfo {author} {\bibfnamefont {I.-Y.}\ \bibnamefont {Oh}}, \bibinfo {author} {\bibfnamefont {M.}~\bibnamefont {Shin}}, \bibinfo {author} {\bibfnamefont {H.}~\bibnamefont {Park}}, \bibinfo {author} {\bibfnamefont {C.}~\bibnamefont {Jang}}, \bibinfo {author} {\bibfnamefont {B.-C.}\ \bibnamefont {Min}}, \bibinfo {author} {\bibfnamefont {S.-I.}\ \bibnamefont {Kim}}, \bibinfo {author} {\bibfnamefont {S.-Y.}\ \bibnamefont {Park}},\ and\ \bibinfo {author} {\bibfnamefont {C.~S.}\ \bibnamefont {Park}},\ }\bibfield  {title} {\bibinfo {title} {Spin nano--oscillator--based wireless communication},\ }\href {https://doi.org/10.1038/srep05486} {\bibfield  {journal} {\bibinfo  {journal} {Scientific Reports}\ }\textbf {\bibinfo {volume} {4}},\ \bibinfo {pages} {5486} (\bibinfo {year} {2014})}\BibitemShut
  {NoStop}%
\bibitem [{\citenamefont {Litvinenko}\ \emph {et~al.}(2020)\citenamefont {Litvinenko}, \citenamefont {Iurchuk}, \citenamefont {Sethi}, \citenamefont {Louis}, \citenamefont {Tyberkevych}, \citenamefont {Li}, \citenamefont {Jenkins}, \citenamefont {Ferreira}, \citenamefont {Dieny}, \citenamefont {Slavin},\ and\ \citenamefont {Ebels}}]{Litvinenko2020}%
  \BibitemOpen
  \bibfield  {author} {\bibinfo {author} {\bibfnamefont {A.}~\bibnamefont {Litvinenko}}, \bibinfo {author} {\bibfnamefont {V.}~\bibnamefont {Iurchuk}}, \bibinfo {author} {\bibfnamefont {P.}~\bibnamefont {Sethi}}, \bibinfo {author} {\bibfnamefont {S.}~\bibnamefont {Louis}}, \bibinfo {author} {\bibfnamefont {V.}~\bibnamefont {Tyberkevych}}, \bibinfo {author} {\bibfnamefont {J.}~\bibnamefont {Li}}, \bibinfo {author} {\bibfnamefont {A.}~\bibnamefont {Jenkins}}, \bibinfo {author} {\bibfnamefont {R.}~\bibnamefont {Ferreira}}, \bibinfo {author} {\bibfnamefont {B.}~\bibnamefont {Dieny}}, \bibinfo {author} {\bibfnamefont {A.}~\bibnamefont {Slavin}},\ and\ \bibinfo {author} {\bibfnamefont {U.}~\bibnamefont {Ebels}},\ }\bibfield  {title} {\bibinfo {title} {Ultrafast sweep-tuned spectrum analyzer with temporal resolution based on a spin-torque nano-oscillator},\ }\href {https://doi.org/10.1021/acs.nanolett.0c02195} {\bibfield  {journal} {\bibinfo  {journal} {Nano Letters}\ }\textbf {\bibinfo {volume} {20}},\ \bibinfo
  {pages} {6104} (\bibinfo {year} {2020})}\BibitemShut {NoStop}%
\bibitem [{\citenamefont {Louis}\ \emph {et~al.}(2017)\citenamefont {Louis}, \citenamefont {Lisenkov}, \citenamefont {Tiberkevich}, \citenamefont {Li}, \citenamefont {Khymyn}, \citenamefont {Bankowksi}, \citenamefont {Meitzler}, \citenamefont {Krivorotov},\ and\ \citenamefont {Slavin}}]{8007619}%
  \BibitemOpen
  \bibfield  {author} {\bibinfo {author} {\bibfnamefont {S.}~\bibnamefont {Louis}}, \bibinfo {author} {\bibfnamefont {I.}~\bibnamefont {Lisenkov}}, \bibinfo {author} {\bibfnamefont {V.}~\bibnamefont {Tiberkevich}}, \bibinfo {author} {\bibfnamefont {J.}~\bibnamefont {Li}}, \bibinfo {author} {\bibfnamefont {R.}~\bibnamefont {Khymyn}}, \bibinfo {author} {\bibfnamefont {E.}~\bibnamefont {Bankowksi}}, \bibinfo {author} {\bibfnamefont {T.}~\bibnamefont {Meitzler}}, \bibinfo {author} {\bibfnamefont {I.}~\bibnamefont {Krivorotov}},\ and\ \bibinfo {author} {\bibfnamefont {A.}~\bibnamefont {Slavin}},\ }\bibfield  {title} {\bibinfo {title} {Low power microwave signal detection with a spin-torque nano-oscillator in the active self-oscillating regime},\ }in\ \href {https://doi.org/10.1109/INTMAG.2017.8007619} {\emph {\bibinfo {booktitle} {2017 IEEE International Magnetics Conference (INTERMAG)}}}\ (\bibinfo {year} {2017})\ pp.\ \bibinfo {pages} {1--1}\BibitemShut {NoStop}%
\bibitem [{\citenamefont {González}\ \emph {et~al.}(2024)\citenamefont {González}, \citenamefont {Litvinenko}, \citenamefont {Kumar}, \citenamefont {Khymyn},\ and\ \citenamefont {Åkerman}}]{GONZALEZ2024101173}%
  \BibitemOpen
  \bibfield  {author} {\bibinfo {author} {\bibfnamefont {V.~H.}\ \bibnamefont {González}}, \bibinfo {author} {\bibfnamefont {A.}~\bibnamefont {Litvinenko}}, \bibinfo {author} {\bibfnamefont {A.}~\bibnamefont {Kumar}}, \bibinfo {author} {\bibfnamefont {R.}~\bibnamefont {Khymyn}},\ and\ \bibinfo {author} {\bibfnamefont {J.}~\bibnamefont {Åkerman}},\ }\bibfield  {title} {\bibinfo {title} {Spintronic devices as next-generation computation accelerators},\ }\href {https://doi.org/https://doi.org/10.1016/j.cossms.2024.101173} {\bibfield  {journal} {\bibinfo  {journal} {Current Opinion in Solid State and Materials Science}\ }\textbf {\bibinfo {volume} {31}},\ \bibinfo {pages} {101173} (\bibinfo {year} {2024})}\BibitemShut {NoStop}%
\bibitem [{\citenamefont {Farkhani}\ \emph {et~al.}(2019)\citenamefont {Farkhani}, \citenamefont {Böhnert}, \citenamefont {Tarequzzaman}, \citenamefont {Costa}, \citenamefont {Jenkins}, \citenamefont {Ferreira},\ and\ \citenamefont {Moradi}}]{8734967}%
  \BibitemOpen
  \bibfield  {author} {\bibinfo {author} {\bibfnamefont {H.}~\bibnamefont {Farkhani}}, \bibinfo {author} {\bibfnamefont {T.}~\bibnamefont {Böhnert}}, \bibinfo {author} {\bibfnamefont {M.}~\bibnamefont {Tarequzzaman}}, \bibinfo {author} {\bibfnamefont {D.}~\bibnamefont {Costa}}, \bibinfo {author} {\bibfnamefont {A.}~\bibnamefont {Jenkins}}, \bibinfo {author} {\bibfnamefont {R.}~\bibnamefont {Ferreira}},\ and\ \bibinfo {author} {\bibfnamefont {F.}~\bibnamefont {Moradi}},\ }\bibfield  {title} {\bibinfo {title} {Spin-torque-nano-oscillator based neuromorphic computing assisted by laser},\ }in\ \href {https://doi.org/10.1109/DTIS.2019.8734967} {\emph {\bibinfo {booktitle} {2019 14th International Conference on Design \& Technology of Integrated Systems In Nanoscale Era (DTIS)}}}\ (\bibinfo {year} {2019})\ pp.\ \bibinfo {pages} {1--5}\BibitemShut {NoStop}%
\bibitem [{\citenamefont {B{\"o}hnert}\ \emph {et~al.}(2023)\citenamefont {B{\"o}hnert}, \citenamefont {Rezaeiyan}, \citenamefont {Claro}, \citenamefont {Benetti}, \citenamefont {Jenkins}, \citenamefont {Farkhani}, \citenamefont {Moradi},\ and\ \citenamefont {Ferreira}}]{Böhnert2023}%
  \BibitemOpen
  \bibfield  {author} {\bibinfo {author} {\bibfnamefont {T.}~\bibnamefont {B{\"o}hnert}}, \bibinfo {author} {\bibfnamefont {Y.}~\bibnamefont {Rezaeiyan}}, \bibinfo {author} {\bibfnamefont {M.~S.}\ \bibnamefont {Claro}}, \bibinfo {author} {\bibfnamefont {L.}~\bibnamefont {Benetti}}, \bibinfo {author} {\bibfnamefont {A.~S.}\ \bibnamefont {Jenkins}}, \bibinfo {author} {\bibfnamefont {H.}~\bibnamefont {Farkhani}}, \bibinfo {author} {\bibfnamefont {F.}~\bibnamefont {Moradi}},\ and\ \bibinfo {author} {\bibfnamefont {R.}~\bibnamefont {Ferreira}},\ }\bibfield  {title} {\bibinfo {title} {Weighted spin torque nano-oscillator system for neuromorphic computing},\ }\href {https://doi.org/10.1038/s44172-023-00117-9} {\bibfield  {journal} {\bibinfo  {journal} {Communications Engineering}\ }\textbf {\bibinfo {volume} {2}},\ \bibinfo {pages} {65} (\bibinfo {year} {2023})}\BibitemShut {NoStop}%
\bibitem [{\citenamefont {Kanao}\ \emph {et~al.}(2019)\citenamefont {Kanao}, \citenamefont {Suto}, \citenamefont {Mizushima}, \citenamefont {Goto}, \citenamefont {Tanamoto},\ and\ \citenamefont {Nagasawa}}]{PhysRevApplied.12.024052}%
  \BibitemOpen
  \bibfield  {author} {\bibinfo {author} {\bibfnamefont {T.}~\bibnamefont {Kanao}}, \bibinfo {author} {\bibfnamefont {H.}~\bibnamefont {Suto}}, \bibinfo {author} {\bibfnamefont {K.}~\bibnamefont {Mizushima}}, \bibinfo {author} {\bibfnamefont {H.}~\bibnamefont {Goto}}, \bibinfo {author} {\bibfnamefont {T.}~\bibnamefont {Tanamoto}},\ and\ \bibinfo {author} {\bibfnamefont {T.}~\bibnamefont {Nagasawa}},\ }\bibfield  {title} {\bibinfo {title} {Reservoir computing on spin-torque oscillator array},\ }\href {https://doi.org/10.1103/PhysRevApplied.12.024052} {\bibfield  {journal} {\bibinfo  {journal} {Phys. Rev. Appl.}\ }\textbf {\bibinfo {volume} {12}},\ \bibinfo {pages} {024052} (\bibinfo {year} {2019})}\BibitemShut {NoStop}%
\bibitem [{\citenamefont {Romera}\ \emph {et~al.}(2018)\citenamefont {Romera}, \citenamefont {Talatchian}, \citenamefont {Tsunegi}, \citenamefont {Abreu~Araujo}, \citenamefont {Cros}, \citenamefont {Bortolotti}, \citenamefont {Trastoy}, \citenamefont {Yakushiji}, \citenamefont {Fukushima}, \citenamefont {Kubota}, \citenamefont {Yuasa}, \citenamefont {Ernoult}, \citenamefont {Vodenicarevic}, \citenamefont {Hirtzlin}, \citenamefont {Locatelli}, \citenamefont {Querlioz},\ and\ \citenamefont {Grollier}}]{Romera2018}%
  \BibitemOpen
  \bibfield  {author} {\bibinfo {author} {\bibfnamefont {M.}~\bibnamefont {Romera}}, \bibinfo {author} {\bibfnamefont {P.}~\bibnamefont {Talatchian}}, \bibinfo {author} {\bibfnamefont {S.}~\bibnamefont {Tsunegi}}, \bibinfo {author} {\bibfnamefont {F.}~\bibnamefont {Abreu~Araujo}}, \bibinfo {author} {\bibfnamefont {V.}~\bibnamefont {Cros}}, \bibinfo {author} {\bibfnamefont {P.}~\bibnamefont {Bortolotti}}, \bibinfo {author} {\bibfnamefont {J.}~\bibnamefont {Trastoy}}, \bibinfo {author} {\bibfnamefont {K.}~\bibnamefont {Yakushiji}}, \bibinfo {author} {\bibfnamefont {A.}~\bibnamefont {Fukushima}}, \bibinfo {author} {\bibfnamefont {H.}~\bibnamefont {Kubota}}, \bibinfo {author} {\bibfnamefont {S.}~\bibnamefont {Yuasa}}, \bibinfo {author} {\bibfnamefont {M.}~\bibnamefont {Ernoult}}, \bibinfo {author} {\bibfnamefont {D.}~\bibnamefont {Vodenicarevic}}, \bibinfo {author} {\bibfnamefont {T.}~\bibnamefont {Hirtzlin}}, \bibinfo {author} {\bibfnamefont {N.}~\bibnamefont {Locatelli}}, \bibinfo {author} {\bibfnamefont
  {D.}~\bibnamefont {Querlioz}},\ and\ \bibinfo {author} {\bibfnamefont {J.}~\bibnamefont {Grollier}},\ }\bibfield  {title} {\bibinfo {title} {Vowel recognition with four coupled spin-torque nano-oscillators},\ }\href {https://doi.org/10.1038/s41586-018-0632-y} {\bibfield  {journal} {\bibinfo  {journal} {Nature}\ }\textbf {\bibinfo {volume} {563}},\ \bibinfo {pages} {230} (\bibinfo {year} {2018})}\BibitemShut {NoStop}%
\bibitem [{\citenamefont {Rippard}\ \emph {et~al.}(2005)\citenamefont {Rippard}, \citenamefont {Pufall}, \citenamefont {Kaka}, \citenamefont {Silva}, \citenamefont {Russek},\ and\ \citenamefont {Katine}}]{PhysRevLett.95.067203}%
  \BibitemOpen
  \bibfield  {author} {\bibinfo {author} {\bibfnamefont {W.~H.}\ \bibnamefont {Rippard}}, \bibinfo {author} {\bibfnamefont {M.~R.}\ \bibnamefont {Pufall}}, \bibinfo {author} {\bibfnamefont {S.}~\bibnamefont {Kaka}}, \bibinfo {author} {\bibfnamefont {T.~J.}\ \bibnamefont {Silva}}, \bibinfo {author} {\bibfnamefont {S.~E.}\ \bibnamefont {Russek}},\ and\ \bibinfo {author} {\bibfnamefont {J.~A.}\ \bibnamefont {Katine}},\ }\bibfield  {title} {\bibinfo {title} {Injection locking and phase control of spin transfer nano-oscillators},\ }\href {https://doi.org/10.1103/PhysRevLett.95.067203} {\bibfield  {journal} {\bibinfo  {journal} {Phys. Rev. Lett.}\ }\textbf {\bibinfo {volume} {95}},\ \bibinfo {pages} {067203} (\bibinfo {year} {2005})}\BibitemShut {NoStop}%
\bibitem [{\citenamefont {Nikitin}\ \emph {et~al.}(2023)\citenamefont {Nikitin}, \citenamefont {Canudas-de Wit}, \citenamefont {Frasca},\ and\ \citenamefont {Ebels}}]{10192065}%
  \BibitemOpen
  \bibfield  {author} {\bibinfo {author} {\bibfnamefont {D.}~\bibnamefont {Nikitin}}, \bibinfo {author} {\bibfnamefont {C.}~\bibnamefont {Canudas-de Wit}}, \bibinfo {author} {\bibfnamefont {P.}~\bibnamefont {Frasca}},\ and\ \bibinfo {author} {\bibfnamefont {U.}~\bibnamefont {Ebels}},\ }\bibfield  {title} {\bibinfo {title} {Synchronization of spin-torque oscillators via continuation method},\ }\href {https://doi.org/10.1109/TAC.2023.3298288} {\bibfield  {journal} {\bibinfo  {journal} {IEEE Transactions on Automatic Control}\ }\textbf {\bibinfo {volume} {68}},\ \bibinfo {pages} {6621} (\bibinfo {year} {2023})}\BibitemShut {NoStop}%
\bibitem [{\citenamefont {Tsunegi}\ \emph {et~al.}(2018)\citenamefont {Tsunegi}, \citenamefont {Taniguchi}, \citenamefont {Lebrun}, \citenamefont {Yakushiji}, \citenamefont {Cros}, \citenamefont {Grollier}, \citenamefont {Fukushima}, \citenamefont {Yuasa},\ and\ \citenamefont {Kubota}}]{Tsunegi2018}%
  \BibitemOpen
  \bibfield  {author} {\bibinfo {author} {\bibfnamefont {S.}~\bibnamefont {Tsunegi}}, \bibinfo {author} {\bibfnamefont {T.}~\bibnamefont {Taniguchi}}, \bibinfo {author} {\bibfnamefont {R.}~\bibnamefont {Lebrun}}, \bibinfo {author} {\bibfnamefont {K.}~\bibnamefont {Yakushiji}}, \bibinfo {author} {\bibfnamefont {V.}~\bibnamefont {Cros}}, \bibinfo {author} {\bibfnamefont {J.}~\bibnamefont {Grollier}}, \bibinfo {author} {\bibfnamefont {A.}~\bibnamefont {Fukushima}}, \bibinfo {author} {\bibfnamefont {S.}~\bibnamefont {Yuasa}},\ and\ \bibinfo {author} {\bibfnamefont {H.}~\bibnamefont {Kubota}},\ }\bibfield  {title} {\bibinfo {title} {Scaling up electrically synchronized spin torque oscillator networks},\ }\href {https://doi.org/10.1038/s41598-018-31769-9} {\bibfield  {journal} {\bibinfo  {journal} {Scientific Reports}\ }\textbf {\bibinfo {volume} {8}},\ \bibinfo {pages} {13475} (\bibinfo {year} {2018})}\BibitemShut {NoStop}%
\bibitem [{\citenamefont {Kumar}\ \emph {et~al.}(2023)\citenamefont {Kumar}, \citenamefont {Litvinenko}, \citenamefont {Behera}, \citenamefont {Awad}, \citenamefont {Khymyn},\ and\ \citenamefont {Åkerman}}]{kumar2023mutualsynchronizationspintorque}%
  \BibitemOpen
  \bibfield  {author} {\bibinfo {author} {\bibfnamefont {A.}~\bibnamefont {Kumar}}, \bibinfo {author} {\bibfnamefont {A.}~\bibnamefont {Litvinenko}}, \bibinfo {author} {\bibfnamefont {N.}~\bibnamefont {Behera}}, \bibinfo {author} {\bibfnamefont {A.~A.}\ \bibnamefont {Awad}}, \bibinfo {author} {\bibfnamefont {R.}~\bibnamefont {Khymyn}},\ and\ \bibinfo {author} {\bibfnamefont {J.}~\bibnamefont {Åkerman}},\ }\href {https://arxiv.org/abs/2312.09656} {\bibinfo {title} {Mutual synchronization in spin torque and spin hall nano-oscillators}} (\bibinfo {year} {2023}),\ \Eprint {https://arxiv.org/abs/2312.09656} {arXiv:2312.09656 [cond-mat.mes-hall]} \BibitemShut {NoStop}%
\bibitem [{\citenamefont {Slavin}\ and\ \citenamefont {Tiberkevich}(2006)}]{PhysRevB.74.104401}%
  \BibitemOpen
  \bibfield  {author} {\bibinfo {author} {\bibfnamefont {A.~N.}\ \bibnamefont {Slavin}}\ and\ \bibinfo {author} {\bibfnamefont {V.~S.}\ \bibnamefont {Tiberkevich}},\ }\bibfield  {title} {\bibinfo {title} {Theory of mutual phase locking of spin-torque nanosized oscillators},\ }\href {https://doi.org/10.1103/PhysRevB.74.104401} {\bibfield  {journal} {\bibinfo  {journal} {Phys. Rev. B}\ }\textbf {\bibinfo {volume} {74}},\ \bibinfo {pages} {104401} (\bibinfo {year} {2006})}\BibitemShut {NoStop}%
\bibitem [{\citenamefont {Mancoff}\ \emph {et~al.}(2005)\citenamefont {Mancoff}, \citenamefont {Rizzo}, \citenamefont {Engel},\ and\ \citenamefont {Tehrani}}]{mancoff2005phase}%
  \BibitemOpen
  \bibfield  {author} {\bibinfo {author} {\bibfnamefont {F.}~\bibnamefont {Mancoff}}, \bibinfo {author} {\bibfnamefont {N.}~\bibnamefont {Rizzo}}, \bibinfo {author} {\bibfnamefont {B.}~\bibnamefont {Engel}},\ and\ \bibinfo {author} {\bibfnamefont {S.}~\bibnamefont {Tehrani}},\ }\bibfield  {title} {\bibinfo {title} {Phase-locking in double-point-contact spin-transfer devices},\ }\href@noop {} {\bibfield  {journal} {\bibinfo  {journal} {Nature}\ }\textbf {\bibinfo {volume} {437}},\ \bibinfo {pages} {393} (\bibinfo {year} {2005})}\BibitemShut {NoStop}%
\bibitem [{\citenamefont {Kaka}\ \emph {et~al.}(2005)\citenamefont {Kaka}, \citenamefont {Pufall}, \citenamefont {Rippard}, \citenamefont {Silva}, \citenamefont {Russek},\ and\ \citenamefont {Katine}}]{kaka2005mutual}%
  \BibitemOpen
  \bibfield  {author} {\bibinfo {author} {\bibfnamefont {S.}~\bibnamefont {Kaka}}, \bibinfo {author} {\bibfnamefont {M.~R.}\ \bibnamefont {Pufall}}, \bibinfo {author} {\bibfnamefont {W.~H.}\ \bibnamefont {Rippard}}, \bibinfo {author} {\bibfnamefont {T.~J.}\ \bibnamefont {Silva}}, \bibinfo {author} {\bibfnamefont {S.~E.}\ \bibnamefont {Russek}},\ and\ \bibinfo {author} {\bibfnamefont {J.~A.}\ \bibnamefont {Katine}},\ }\bibfield  {title} {\bibinfo {title} {Mutual phase-locking of microwave spin torque nano-oscillators},\ }\href@noop {} {\bibfield  {journal} {\bibinfo  {journal} {Nature}\ }\textbf {\bibinfo {volume} {437}},\ \bibinfo {pages} {389} (\bibinfo {year} {2005})}\BibitemShut {NoStop}%
\bibitem [{\citenamefont {Castro}\ \emph {et~al.}(2022)\citenamefont {Castro}, \citenamefont {Mancilla-Almonacid}, \citenamefont {Dieny}, \citenamefont {Allende}, \citenamefont {Buda-Prejbeanu},\ and\ \citenamefont {Ebels}}]{Castro2022}%
  \BibitemOpen
  \bibfield  {author} {\bibinfo {author} {\bibfnamefont {M.~A.}\ \bibnamefont {Castro}}, \bibinfo {author} {\bibfnamefont {D.}~\bibnamefont {Mancilla-Almonacid}}, \bibinfo {author} {\bibfnamefont {B.}~\bibnamefont {Dieny}}, \bibinfo {author} {\bibfnamefont {S.}~\bibnamefont {Allende}}, \bibinfo {author} {\bibfnamefont {L.~D.}\ \bibnamefont {Buda-Prejbeanu}},\ and\ \bibinfo {author} {\bibfnamefont {U.}~\bibnamefont {Ebels}},\ }\bibfield  {title} {\bibinfo {title} {Mutual synchronization of spin-torque oscillators within a ring array},\ }\href {https://doi.org/10.1038/s41598-022-15483-1} {\bibfield  {journal} {\bibinfo  {journal} {Scientific Reports}\ }\textbf {\bibinfo {volume} {12}},\ \bibinfo {pages} {12030} (\bibinfo {year} {2022})}\BibitemShut {NoStop}%
\bibitem [{\citenamefont {Prabhakar}\ and\ \citenamefont {Stancil}(2009)}]{prabhakar2009spin}%
  \BibitemOpen
  \bibfield  {author} {\bibinfo {author} {\bibfnamefont {A.}~\bibnamefont {Prabhakar}}\ and\ \bibinfo {author} {\bibfnamefont {D.~D.}\ \bibnamefont {Stancil}},\ }\href@noop {} {\emph {\bibinfo {title} {Spin waves: Theory and applications}}},\ Vol.~\bibinfo {volume} {5}\ (\bibinfo  {publisher} {Springer},\ \bibinfo {year} {2009})\BibitemShut {NoStop}%
\bibitem [{\citenamefont {Fassbender}\ and\ \citenamefont {McCord}(2008)}]{FASSBENDER2008579}%
  \BibitemOpen
  \bibfield  {author} {\bibinfo {author} {\bibfnamefont {J.}~\bibnamefont {Fassbender}}\ and\ \bibinfo {author} {\bibfnamefont {J.}~\bibnamefont {McCord}},\ }\bibfield  {title} {\bibinfo {title} {Magnetic patterning by means of ion irradiation and implantation},\ }\href {https://doi.org/https://doi.org/10.1016/j.jmmm.2007.07.032} {\bibfield  {journal} {\bibinfo  {journal} {Journal of Magnetism and Magnetic Materials}\ }\textbf {\bibinfo {volume} {320}},\ \bibinfo {pages} {579} (\bibinfo {year} {2008})}\BibitemShut {NoStop}%
\bibitem [{\citenamefont {Wimmer}\ \emph {et~al.}(2019)\citenamefont {Wimmer}, \citenamefont {Althammer}, \citenamefont {Liensberger}, \citenamefont {Vlietstra}, \citenamefont {Gepr\"ags}, \citenamefont {Weiler}, \citenamefont {Gross},\ and\ \citenamefont {Huebl}}]{PhysRevLett.123.257201}%
  \BibitemOpen
  \bibfield  {author} {\bibinfo {author} {\bibfnamefont {T.}~\bibnamefont {Wimmer}}, \bibinfo {author} {\bibfnamefont {M.}~\bibnamefont {Althammer}}, \bibinfo {author} {\bibfnamefont {L.}~\bibnamefont {Liensberger}}, \bibinfo {author} {\bibfnamefont {N.}~\bibnamefont {Vlietstra}}, \bibinfo {author} {\bibfnamefont {S.}~\bibnamefont {Gepr\"ags}}, \bibinfo {author} {\bibfnamefont {M.}~\bibnamefont {Weiler}}, \bibinfo {author} {\bibfnamefont {R.}~\bibnamefont {Gross}},\ and\ \bibinfo {author} {\bibfnamefont {H.}~\bibnamefont {Huebl}},\ }\bibfield  {title} {\bibinfo {title} {Spin transport in a magnetic insulator with zero effective damping},\ }\href {https://doi.org/10.1103/PhysRevLett.123.257201} {\bibfield  {journal} {\bibinfo  {journal} {Phys. Rev. Lett.}\ }\textbf {\bibinfo {volume} {123}},\ \bibinfo {pages} {257201} (\bibinfo {year} {2019})}\BibitemShut {NoStop}%
\bibitem [{\citenamefont {Diebner}\ \emph {et~al.}(2002)\citenamefont {Diebner}, \citenamefont {Sahle},\ and\ \citenamefont {Hoff}}]{DIEBNER2002781}%
  \BibitemOpen
  \bibfield  {author} {\bibinfo {author} {\bibfnamefont {H.~H.}\ \bibnamefont {Diebner}}, \bibinfo {author} {\bibfnamefont {S.}~\bibnamefont {Sahle}},\ and\ \bibinfo {author} {\bibfnamefont {A.~A.}\ \bibnamefont {Hoff}},\ }\bibfield  {title} {\bibinfo {title} {A realtime adaptive system for dynamics recognition},\ }\href {https://doi.org/https://doi.org/10.1016/S0960-0779(01)00053-4} {\bibfield  {journal} {\bibinfo  {journal} {Chaos, Solitons \& Fractals}\ }\textbf {\bibinfo {volume} {13}},\ \bibinfo {pages} {781} (\bibinfo {year} {2002})}\BibitemShut {NoStop}%
\bibitem [{\citenamefont {Boerner}\ \emph {et~al.}(2023)\citenamefont {Boerner}, \citenamefont {Deems}, \citenamefont {Furlani}, \citenamefont {Knuth},\ and\ \citenamefont {Towns}}]{10.1145/3569951.3597559}%
  \BibitemOpen
  \bibfield  {author} {\bibinfo {author} {\bibfnamefont {T.~J.}\ \bibnamefont {Boerner}}, \bibinfo {author} {\bibfnamefont {S.}~\bibnamefont {Deems}}, \bibinfo {author} {\bibfnamefont {T.~R.}\ \bibnamefont {Furlani}}, \bibinfo {author} {\bibfnamefont {S.~L.}\ \bibnamefont {Knuth}},\ and\ \bibinfo {author} {\bibfnamefont {J.}~\bibnamefont {Towns}},\ }\bibfield  {title} {\bibinfo {title} {Access: Advancing innovation: Nsf’s advanced cyberinfrastructure coordination ecosystem: Services \& support},\ }in\ \href {https://doi.org/10.1145/3569951.3597559} {\emph {\bibinfo {booktitle} {Practice and Experience in Advanced Research Computing}}},\ \bibinfo {series and number} {PEARC '23}\ (\bibinfo  {publisher} {Association for Computing Machinery},\ \bibinfo {address} {New York, NY, USA},\ \bibinfo {year} {2023})\ p.\ \bibinfo {pages} {173–176}\BibitemShut {NoStop}%
\end{thebibliography}%

\end{document}